\documentclass[journal,comsoc]{IEEEtran}
\IEEEoverridecommandlockouts
\usepackage{cite}
\usepackage{amsmath,amssymb,amsfonts}
\usepackage{algorithmic}
\usepackage{graphicx}
\usepackage{textcomp}
\usepackage{xcolor}

\usepackage[T1]{fontenc}

\usepackage[hyphens]{url}
\usepackage{hyperref}
\hypersetup{breaklinks=true}

\graphicspath{{myfigures/}}
\usepackage{epstopdf}					
\usepackage{amsfonts}
\usepackage{amssymb}
\usepackage{multirow}
\usepackage{bm}
\usepackage{cite}

\newtheorem{theorem}{Theorem}[section]
\newcommand{\nid}{\noindent}

\usepackage{color}

\ifCLASSINFOpdf
\else
\fi
\usepackage{amsmath}
\usepackage[cmintegrals]{newtxmath}
\begin{document}
%
\title{Low-Density Spreading Design Based on an Algebraic Scheme for NOMA Systems}
%
%
%

\author{Goldwyn~Millar, Michel~Kulhandjian,~\IEEEmembership{Senior Member,~IEEE,
}Ayse~Alaca, Saban~Alaca,  Claude~D'Amours,~\IEEEmembership{Member,~IEEE,
}and Halim~Yanikomeroglu,~\IEEEmembership{Fellow,~IEEE
}

\thanks{M. Kulhandjian and C. D'Amours are with the School of Electrical Engineering, \& Computer Science, University of Ottawa, Ottawa, Canada, e-mail: mkk6@buffalo.edu, cdamours@uottawa.ca.}
\thanks{G. Millar, A. Alaca and S. Alaca are with the School of Mathematics and Statistics, Carleton University, Ottawa, Canada, e-mail:
goldwynmillar@cmail.carleton.ca, \{aysealaca,sabanalaca\}@cunet.carleton.ca.}
\thanks{H. Yanikomeroglu is with the Department of Systems \& Computer Engineering, Carleton University, Ottawa, Canada, e-mail: halim@sce.carleton.ca.}
\thanks{Manuscript received Dec. 03, 2021;}
}

%
%

\markboth{IEEE Wireless Communications Letters,~Vol.~1, No.~1, Dec.~2021}%
{Shell \MakeLowercase{\textit{et al.}}: Bare Demo of IEEEtran.cls for IEEE Communications Society Journals}
%



\maketitle

\begin{abstract}
In this paper, a code-domain nonorthogonal
multiple access (NOMA) technique based on an algebraic design is studied. We propose an improved low-density spreading (LDS) sequence design based on projective geometry. In terms of its bit error rate (BER) performance, our proposed improved LDS code set outperforms the existing LDS designs over the frequency-nonselective Rayleigh fading and additive white Gaussian noise (AWGN) channels. We demonstrated that achieving the best BER depends on the minimum distance.  

\end{abstract}


%
\IEEEpeerreviewmaketitle

\section{Introduction}
%
%
%
%

\IEEEPARstart{I}{n} previous generations of wireless communications, orthogonal multiple access (OMA) techniques were predominant. In OMA systems, users are assigned resources that are orthogonal to one another, such as orthogonal frequency division multiple access (OFDMA), orthogonal time division multiple access (OTDMA) or code division multiple access (CDMA) where the spreading signatures were mutually orthogonal. Ideally, in OMA systems the presence of multiple users does not cause interference to any of the users that occupy the channel. However, the capacity of an OMA system is limited by the number of available orthogonal resources \cite{Dai2018}.

Future wireless networks are required to support a wide range of use cases. One such use case is to provide communication capabilities to a massive number of low-power Internet-of-Things (IoT) devices \cite{Dai2018}. Due to the limitations of OMA, supporting a large number of users over a common channel while achieving the required level of service quality may not be possible. In rank-deficient cases, where the number of active communication devices exceeds the number of orthogonal resources, nonorthogonal multiple access (NOMA) systems are proposed \cite{Dai2018}. In NOMA systems, users are assigned multiple orthogonal resources and the resources are assigned to multiple users. When users transmit simultaneously, each user may experience some multiple-access interference (MAI) if the same resource is used by two or more transmitters simultaneously. For mitigating the MAI, many researchers have proposed a sparse allocation of these resources so as to take advantage of efficient sparse signal processing techniques. For example, the message passing algorithm (MPA) can be used to iteratively perform multiuser detection (MUD) in a NOMA system using sparsely assigned resources. NOMA techniques can be categorized into power-domain multiplexed NOMA (PDM-NOMA) and code domain multiplexed NOMA (CDM-NOMA) \cite{Dai2018}. A few of the strong contenders of CDM-NOMA are low-density spread CDMA (LDS-CDMA) \cite{Hoshyar2008}, low-density spread orthogonal frequency-division multiplexing (LDS-OFDM) \cite{Razavi2012}, and sparse code multiple access (SCMA) \cite{Nikopour2013}.

A number of studies have been undertaken to design spreading code sets for sparse spreading based NOMA \cite{Hoshyar2006, Song2017, Jiang2019}. In \cite{Hoshyar2006} the authors proposed an LDS structure based on LDPC codes, where the user's symbols are arranged in such a way that the interference seen by each user on each chip is different, while in \cite{JVan2009}, the authors designed the spreading sequences based on an LDPC indicator matrix. In general, signatures having a unity scalar magnitude are designed by maximizing their minimum Euclidean distance. Similar to the minimum distance criterion based LDS code design of \cite{JVan2009}, the authors of \cite{Song2017} consider the maximization of the minimum Euclidean distance for QAM constellations. Notably, they design signature matrices that have factor graphs exhibiting very few short cycles and large superposed signal constellation distances. In \cite{Jiang2019}, the authors optimize the degree distribution of the LDS signature matrix.

Combinatorial structures with balanced incomplete block design (BIBD) can allow a large number of users employing few resources. As an example, authors in \cite{Claude1995} proposed a multiple tone frequency shift keying (MT-FSK) waveform design based on BIBD. A specific highly structured BIBD called the Steiner triple system (STS) is well studied for low density parity-check (LDPC) constructions \cite{Ivanov2013}. Steiner designs used as sparse codes have better interference properties, which provide higher user/bandwidth efficiency and variable code rates. Motivated by this, Wu \emph{et al.} \cite{Atkin2018} proposed a STS-based LDS signature set design, whose incidence matrix supports superposition based multiuser communications. By using algebraic code construction methods, the authors in \cite{Mheich2019, Xudong2021} proposed a power-imbalanced LDS design of nonzero entries for a given factor graph with the aid of Eisenstein integers\footnote{Eisenstein integers are complex number of the form $z = a+b \omega$, where $a,b \in \mathbb{N}$ and $\omega = \frac{-1 + i\sqrt{3}}{2} = e^{\frac{2\pi i}{3}}$ is a primitive cube root of unity.}. Compared to the design of conventional antipodal spreading sequences for classic CDMA, designing the LDS sequences for NOMA systems is more complicated, since the design should be implemented under the sparsity constraint of the signature matrix. In the literature, there is a little work on the design of optimal signature matrix designs that maximize the minimum Euclidean distance. 

In this paper, we study an LDS design constructed using lines and quadrics from certain finite projective planes. More explicitly, our new contributions are summarized as follows:
      \vspace{-0.0cm}
      \begin{enumerate}
      \item We propose a novel LDS design based on algebraic scheme. Explicitly, our design constructs the incidence matrices by applying Singer's Theorem.   
       \item We demonstrate that our proposed code sets achieve TSC asymptotically by comparing with the widely known Welch bound. Furthermore, we provide a proof that the maximum minimum Euclidean distance of the column vectors is $\sqrt{2}$.
      \end{enumerate}

The rest of the paper is organized as follows. In Section \ref{sec:prelimiaries}, we discuss the preliminaries. In Section \ref{lines and quadrics}, we introduce the facts from projective geometry that are required for our construction. We provide the actual design of our spreading codes in Section \ref{proposedLDS}, along with a theoretical analysis of the properties of our code matrices. After illustrating our simulation results in Section \ref{simulation}, our conclusions are drawn in Section \ref{conclusion}.
\section{System Model}
First of all, perfect chip synchronization among all the transmitters is assumed. In our multiple-access system the users' symbols are multiplexed after spreading them using the LDS codes. Mathematically, we can formulate the system model as\vspace{-0.0cm}
		\begin{eqnarray}
	\label{systemModelAWGN} \mathbf{y} &=& \sum_{k = 1}^K \mathbf{c}_k d_k x_k + \mathbf{n} \nonumber \\  &=& \mathbf{C}\mathbf{D}\mathbf{x} + \mathbf{n},
	\end{eqnarray}
	\nid where $K$ is the number of the users, $d_k$ is the $k$-th user's amplitude, $x_k \in \mathcal{X}_k$ is the $k$-th user's symbol to be transmitted from the constellation alphabet, $\mathcal{X}_k$, $\mathbf{C}= [\mathbf{c}_1, \mathbf{c}_2, \dots, \mathbf{c}_K] \in \mathbb{C}^{L \times K}$ is the column-normalized LDS code matrix, $||\mathbf{c}_k|| = 1$ for $1 \leq k \leq K$, $\mathbf{n}\in \mathbb{C}^{L \times 1}$ is an $L$-dimensional complex-valued AWGN vector with variance of $\sigma^2$ and $\mathbf{D}$ is a diagonal matrix hosting the users' amplitudes. We assume that the constellation alphabet of each user is identical, i.e., $\mathcal{X}_k = \mathcal{X}$, $\forall k$ and the cardinality of the constellation is $M = | \mathcal{X} |$.

\vspace{-0.2cm}
\section{Preliminaries}
\label{sec:prelimiaries}
\subsection{Desiderata}

We are interested in ``overloaded'' spreading matrices, i.e., matrices for which $K > L$. Additionally, in order to achieve good LDS spreading sets, we would like for the maximum cross-correlation and the total squared correlation of our matrices to be low; ideally, these quantities should be as close to the Welch bounds as possible.
\vspace{-0.0cm}
\section{Background from projective geometry}
\label{lines and quadrics}
\subsection{Definitions and Basic Facts}

Let $\mathcal{P}$ be a finite projective plane of order $k$. Enumerate the points and lines of $\mathcal{P}$ using the integers from the set $\lbrace 0,1,...,k^2+k \rbrace$. Then the \emph{incidence matrix} of $\mathcal{P}$ relative to this enumeration is the $(k^2+k+1)\times(k^2+k+1)$ matrix $B = [b_{ij}]$ such that $b_{ij}=1$ if point $i$ is on line $j$ and $b_{ij}=0$ otherwise.

The smallest nontrivial projective plane is the projective plane of order $2$, which is also known as the Fano plane. Relative to a certain enumeration of its points and lines, the incidence matrix for the Fano plane is as follows:
\begin{equation}
\mathbf{I}_{7}=\begin{bmatrix} 0 & 0 & 0 & 1 & 0 & 1 & 1 \\
1 & 0 & 0 & 0 & 1 & 0 & 1 \\
1 & 1 & 0 & 0 & 0 & 1 & 0 \\
0 & 1 & 1 & 0 & 0 & 0 & 1 \\
1 & 0 & 1 & 1 & 0 & 0 & 0 \\
0 & 1 & 0 & 1 & 1 & 0 & 0 \\
0 & 0 & 1 & 0 & 1 & 1 & 0 \\
\end{bmatrix}.
\end{equation}
Incidence matrices of projective planes will be our starting point for our construction of overloaded code matrices for LDS. 

\subsection{Singer's Theorem}
\label{Singer}
In this section, we discuss a well-known theorem first proven by J. Singer in 1938 \cite{Singer1938}.
\begin{theorem} \cite{Singer1938} Let $q$ be a prime power, let $V = \mathbb{F}_q^3,$ and let $\alpha$ be a generator of $\mathbb{F}_{q^3}$. Then there exists a labelling of $P(V)$ such that the resulting incidence matrix is circulant. The first column of this matrix is obtained as follows: we set the entry in row $i$, col $0$ to be $1$ if $\text{Tr}(\alpha^i)= 0$ and to be $0$ otherwise.
\end{theorem}

\begin{equation}
\setcounter{MaxMatrixCols}{20}
\mathbf{I}_{13}= \begin{bmatrix} 1 & 0 & 0 & 0 & 1 & 0 & 0 & 0 & 0 & 0 & 1 & 0 & 1 \\
1 & 1 & 0 & 0 & 0 & 1 & 0 & 0 & 0 & 0 & 0 & 1 & 0 \\
0 & 1 & 1 & 0 & 0 & 0 & 1 & 0 & 0 & 0 & 0 & 0 & 1 \\
1 & 0 & 1 & 1 & 0 & 0 & 0 & 1 & 0 & 0 & 0 & 0 & 0 \\
0 & 1 & 0 & 1 & 1 & 0 & 0 & 0 & 1 & 0 & 0 & 0 & 0 \\
0 & 0 & 1 & 0 & 1 & 1 & 0 & 0 & 0 & 1 & 0 & 0 & 0 \\
0 & 0 & 0 & 1 & 0 & 1 & 1 & 0 & 0 & 0 & 1 & 0 & 0 \\
0 & 0 & 0 & 0 & 1 & 0 & 1 & 1 & 0 & 0 & 0 & 1 & 0 \\
0 & 0 & 0 & 0 & 0 & 1 & 0 & 1 & 1 & 0 & 0 & 0 & 1 \\
1 & 0 & 0 & 0 & 0 & 0 & 1 & 0 & 1 & 1 & 0 & 0 & 0 \\
0 & 1 & 0 & 0 & 0 & 0 & 0 & 1 & 0 & 1 & 1 & 0 & 0 \\
0 & 0 & 1 & 0 & 0 & 0 & 0 & 0 & 1 & 0 & 1 & 1 & 0\\
0 & 0 & 0 & 1 & 0 & 0 & 0 & 0 & 0 & 1 & 0 & 1 & 1
\end{bmatrix} \label{I13}
\end{equation}
We will use Singer's Theorem to construct the incidence matrix of $P(\mathbb{F}_3^3)$. Note that $x^3+2x+1$ is irreducible over $\mathbb{F}_3.$ and therefore the Galois field $\mathbb{F}_{3^3}$ can be constructed as $\mathbb{F}_{3^3} = \mathbb{F}_3[x]/\langle x^3+2x+1 \rangle$. In fact, it turns out that $x^3+2x+1$ is a primitive polynomial over $\mathbb{F}_3,$ which means that $x+\langle x^3+2x+1\rangle$ is a generator of $\mathbb{F}_{3^3}^*$ (i.e., $\alpha = x + \langle x^3+2x+1 \rangle$). The cyclic group $\mathbb{F}_{3^3}^*/\mathbb{F}_{3}^*$ induces an automorphism group of $P(\mathbb{F}_3^3)$ which acts sharply transitively on points and hyperplanes \cite{Arasu1995}.

In order to use Singer's theorem to construct our incidence matrix, we need to be able to compute $\text{Tr}(\alpha^i)$ for $0 \leq i \leq (3^2+3),$ i.e., $0 \leq i \leq 12$, as described in \cite{Arasu1995}. This can be accomplished by first constructing a dictionary between the additive and multiplicative representations of the elements of $\mathbb{F}_{27}^*.$ Here are the values we found for the trace function: $\text{Tr}(\alpha^0) = 0,$ $\text{Tr}(\alpha^1) = 0,$ $\text{Tr}(\alpha^2) = 2,$ $\text{Tr}(\alpha^3) = 0,$ $\text{Tr}(\alpha^4) = 2,$ $\text{Tr}(\alpha^5) = 1,$ $\text{Tr}(\alpha^6) = 2,$ $\text{Tr}(\alpha^7) = 2,$ $\text{Tr}(\alpha^8) = 1,$ $\text{Tr}(\alpha^9) = 0,$ $\text{Tr}(\alpha^{10}) = 2,$ $\text{Tr}(\alpha^{11}) = 2,$ and $\text{Tr}(\alpha^{12}) = 2$. Therefore, the incidence matrix for $P(\mathbb{F}_3^3)$ is as shown in (\ref{I13}).
\subsection{Quadrics}
\label{quadrics}

Let $Q$ be a non-degenerate quadric in $P(\mathbb{F}_q^3)$. Define the incidence vector $\mathbf{g}_Q$ of $Q$ as follows. Let the entry in the $i$-th row equal $1$ if $\alpha^i \in Q$ and $0$ if $\alpha^i \not\in Q$.
\begin{theorem} \cite{Arasu1995} Let $q$ be a prime power, and let $r \in (\mathbb{Z}/(q^2+q+1)\mathbb{Z})^*.$ If $q$ is even, let $r$ be such that $r^{-1} = q+1$; if $q$ is odd, let $r$ be such that $r^{-1} = 2$. Then there exists a non-degenerate quadric $Q$ in $P(\mathbb{F}_q^3)$ whose incidence vector $\mathbf{g}_Q$ is obtained as follows. Let $\mathbf{g}_{\ell}$ be the first column of the incidence matrix of $P(\mathbb{F}_q^3)$ (obtained using Singer's construction). Then the $ri$-th row entry of $\mathbf{g}_Q$ (where $ri$ is reduced modulo $q^2+q+1$) is the same as the $i$-th row entry of $\mathbf{g}_{\ell}$.
\end{theorem} 

Let $q=2$, we have $r^{-1} = 2+1 = 3$ where $r = 3^{-1} = 5$. The first column of $\mathbf{I}_7$ is $[0, 1, 1, 0, 1, 0, 0]^T$. Hence, $\mathbf{g}_Q = [ 0,0,0,1,0,1,1]^T$.
When $q = 3$, we have $r^{-1}=2$ where $r = 2^{-1} = 7$. The first column in the incidence matrix of $P(\mathbb{F}_3^3)$ we derived earlier is $[1,1,0,1,0,0,0,0,0,1,0,0,0]^T$. So, $\mathbf{g}_Q = [ 1,0,0,0,0,0,0,1,1,0,0,1,0]^T$.
. 
\begin{theorem} \label{intersectthm} Let $q$ be a prime power, and let $Q$ be a non-degenerate quadric in $P(\mathbb{F}_q^3)$. Then the lines of $P(\mathbb{F}_q^3)$ intersect $Q$ in sets of sizes $0,$ $1,$ and $2$ and with multiplicities $A,$ $B,$ and $C$, respectively, where 
\[A = \frac{q^2-q}{2}, \hspace{0.1in} B = q+1, \hspace{0.1in} \text{and} \hspace{0.1in} C = \frac{q^2+q}{2}.\]
\end{theorem}
\vspace{-0.4cm}
\section{Proposed LDS code design}
\label{proposedLDS}
\subsection{LDS Construction}
\label{contruction}
In the following section, we describe the proposed algorithm as shown in Table \ref{LDSAlgebraicDesign}. 
    \begin{table}[h]
		\vspace{-0.0cm} \caption {}
		\centering  %
		\begin{tabular}{l}
			\hline \hline \rule{0pt}{3ex} 
			\nid \textbf{LDS design algorithm}  \\
			\hline \rule{0pt}{3ex} 
			\nid \textbf{{Input}:} $L$;  \\
			\hspace{0.1cm} 1: \hspace{0.0cm} Construct incidence matrix, $\mathbf{I}_L$ \\
			\hspace{0.1cm} 2: \hspace{0.0cm} Compute $\mathbf{g}_Q$  \\ 
			\hspace{0.1cm} 3: \hspace{0.0cm}  Generate vector $\mathbf{g}_Q^{\prime}$ from $\mathbf{g}_Q$ \\
			\hspace{0.1cm} 4: \hspace{0.0cm} $\mathbf{C}\gets [\mathbf{I}_L \: \mathbf{g}_Q \: \mathbf{g}_Q^{\prime}]$ \\ 
			\hspace{0.1cm} 5:  \hspace{0.0cm} Negate and normalize $ \mathbf{C}$ \\
			\nid \textbf{{Output}:} $\mathbf{{C}}$ \\
			\hline
		\end{tabular}\vspace{-0.0cm}
		\label{LDSAlgebraicDesign}
	\end{table}
	
	\vspace{-0.0cm}

First, we generate the incidence matrix $\mathbf{I}_L$ of the projective plane $P(\mathbb{F}_q^3)$. Next, append $\mathbf{g}_Q$ vector to incidence matrix $\mathbf{I}_L$. Now, identify every line $\ell$ that intersects $Q$ in two points. Then, in the columns  corresponding to such lines, we negate one of the entries corresponding to either of the points of intersection. Perform a downward cyclic shift on $\mathbf{g}_Q$ to obtain a vector $\mathbf{g}_Q^{\prime}$ that has a suitably small number of nonzero entries in common with $\mathbf{g}_Q$. It follows from the circulant structure of the incidence matrix $\mathbf{I}_L$ that the dot product of $\mathbf{g}_Q^{\prime}$ with the columns of the $\mathbf{I}_L$ also equal either $0$, $1$, or $2$. The next step is to negate entries of the incidence matrix so that the dot product of $\mathbf{g}_Q^{\prime}$ with its columns is less than or equal to $1$. For example, the LDS code construction of size of $7\times 9$ are given as follows: 
\[ \begin{bmatrix} 0 & 0 & 0 & 1 & 0 & 1 & 1 & 0\\
1 & 0 & 0 & 0 & 1 & 0 & 1 & 0\\
1 & 1 & 0 & 0 & 0 & 1 & 0 & 0\\
0 & -1 & -1 & 0 & 0 & 0 & 1 & 1\\
1 & 0 & 1 & 1 & 0 & 0 & 0 & 0\\
0 & 1 & 0 & 1 & -1 & 0 & 0 & 1\\
0 & 0 & 1 & 0 & 1 & 1 & 0 & 1 \end{bmatrix}. \]
Next, we obtain a downward cyclic shift (by one position) on $\mathbf{g}_Q$ to obtain the vector $\mathbf{g}_Q^{\prime} = [1,0,0,0,1,0,1]^T$. 
Note that $\mathbf{g}_Q \cdot \mathbf{g}_Q^{\prime} = 1.$ In the last step of the construction process adjoin $\mathbf{g}_Q^{\prime}$ to the matrix, negating entries so that the dot product of any two columns is less than or equal to $1$, and normalize all of the columns. Therefore, LDS code of size $7\times 9$ is obtained as follows:
\[ (1/\sqrt{3} )\begin{bmatrix} 0 & 0 & 0 & 1 & 0 & 1 & 1 & 0 & 1\\
1 & 0 & 0 & 0 & 1 & 0 & 1 & 0 & 0\\
1 & 1 & 0 & 0 & 0 & 1 & 0 & 0 & 0\\
0 & -1 & -1 & 0 & 0 & 0 & 1 & 1 & 0\\
1 & 0 & -1 & -1 & 0 & 0 & 0 & 0 & 1\\
0 & 1 & 0 & 1 & -1 & 0 & 0 & 1 & 0\\
0 & 0 & 1 & 0 & 1 & -1 & 0 & 1 & 1 \end{bmatrix}. \]
Using our proposed algorithm in Table \ref{LDSAlgebraicDesign}, we can construct LDS code set having size of $13 \times 15$ as follows:
\setcounter{MaxMatrixCols}{20}
\[ (\tiny 1/2)\begin{bmatrix} 1 & 0 & 0 & 0 & 1 & 0 & 0 & 0 & 0 & 0 & 1 & 0 & 1 & 1 & 0 \\
1 & 1 & 0 & 0 & 0 & 1 & 0 & 0 & 0 & 0 & 0 & 1 & 0 & 0 & 1 \\
0 & 1 & 1 & 0 & 0 & 0 & 1 & 0 & 0 & 0 & 0 & 0 & 1 & 0 & 0 \\
1 & 0 & 1 & 1 & 0 & 0 & 0 & 1 & 0 & 0 & 0 & 0 & 0 & 0 & 0 \\
0 & 1 & 0 & 1 & 1 & 0 & 0 & 0 & 1 & 0 & 0 & 0 & 0 & 0 & 0 \\
0 & 0 & 1 & 0 & 1 & 1 & 0 & 0 & 0 & 1 & 0 & 0 & 0 & 0 & 0 \\
0 & 0 & 0 & 1 & 0 & 1 & 1 & 0 & 0 & 0 & 1 & 0 & 0 & 0 & 0 \\
0 & 0 & 0 & 0 & -1 & 0 & 1 & -1 & 0 & 0 & 0 & 1 & 0 & 1 & 0 \\
0 & 0 & 0 & 0 & 0 & -1 & 0 & 1 & -1 & 0 & 0 & 0 & -1 & 1 & 1 \\
-1 & 0 & 0 & 0 & 0 & 0 & 1 & 0 & 1 & -1 & 0 & 0 & 0 & 0 & 1 \\
0 & 1 & 0 & 0 & 0 & 0 & 0 & 1 & 0 & 1 & 1 & 0 & 0 & 0 & 0 \\
0 & 0 & 1 & 0 & 0 & 0 & 0 & 0 & 1 & 0 & -1 & -1 & 0 & 1 & 0 \\
0 & 0 & 0 & 1 & 0 & 0 & 0 & 0 & 0 & 1 & 0 & -1 & 1 & 0 & 1
\end{bmatrix}. \]

\subsection{Analysis}
\label{Analysis}
Let us consider the $(q^2+q+1) \times (q^2+q+2)$ matrices obtained by performing the first stage of the process outlined in Section \ref{contruction}. As for our future LDS code design development, we are working on deriving some theoretical results for the code sets having greater number of users by performing the second step (or, even by performing further iterations) of our proposed construction process. Therefore, the theoretical results of such overloaded matrices will be necessarily built upon our analysis of proposed code matrices that have lower number of users.

Consider a $(q^2+q+1)\times (q^2+q+2)$ matrix obtained by performing the first stage of the procedure outlined in \ref{Singer}. Since any two lines in a projective plane intersect one another in exactly $1$ point, it follows that the cross-correlation of any two columns indexing lines in the plane equals $\pm 1/(q+1).$ Due to the fact that a quadric intersects a line in $0$, $1$, or $2$ points and the way we assign signs to some of the entries in our construction, it follows that the cross-correlation of $\mathbf{g}_Q$ with a column indexing a line is less than or equal to $1/(q+1)$ in absolute value. Hence, the maximum cross-correlation of our code matrix is $1/(q+1)$. Therefore, relative to the size of the alphabet used to construct our sequences and the number of nonzero entries appearing in each row and column of the matrix, the maximum cross-correlation of our code matrix is \emph{optimal}. The vectors in our code matrix have length $q^2+q+1$ and contain only $q+1$ nonzero entries. The larger $q$ results in more sparseness in our proposed vectors.

To consider the total squared (TSC) correlation criteria, we examine how close our code matrices come to the Welch bound, 
\[\textsf{TSC} \geq \frac{K^2}{L} = \frac{(q^2+q+2)^2}{q^2+q+1} = \mathcal{O}(q^2).\]
Since our vectors have unit length, the correlations of the vectors with themselves contribute $q^2+q+1$ to the TSC. Consider the contributions to the TSC provided by pairs of columns corresponding to lines. Any two lines intersect in exactly one point. 

Finally, consider the contribution of the correlations of $\mathbf{g}_Q$ with the columns indexing lines. For the columns indexing lines that do not intersect $Q$ or that intersect $Q$ in $2$ points, the contribution of this cross-correlation to the TSC is $0$. By Theorem \ref{intersectthm}, there are exactly $q+1$ columns indexing lines that intersect $Q$ in $1$ point. The cross-correlations of $\mathbf{g}_Q$ with these lines contribute exactly $(q+1) \cdot {1}/{(q+1)^2} = {1}/{(q+1)}$
to the TSC. Hence, 
\[\textsf{TSC} = (q^2+q+2) + \frac{(q^2+q+1)(q^2+q)}{2(q+1)^2} + \frac{1}{q+1}\] 
Therefore, TSC of our code matrices asymptotically equal to the Welch bound.

It is clear from the discussion about cross-correlation given above that the Hamming distance between any two vectors in the $(q^2+q+1)\times (q^2+q+2)$ code matrix constructed using our procedure is either $2(q+1)$, $2q$, or $2(q-1)$. When the Hamming distance is $2(q+1)$, the Euclidean distance is $\sqrt{{2(q+1)}/{(q+1)}} = \sqrt{2}$.
When the Hamming distance is $2q$, the Euclidean distance is either $\sqrt{{2q}/{(q+1)}}$ or $\sqrt{{(2q+4)}/{(q+1)}}$.
Note we have signed certain elements in the columns, when the Hamming distance is $2(q-1)$, the Euclidean distance is 
\[\sqrt{\frac{2(q-1)+4}{q+1}} = \sqrt{\frac{2(q+1)}{q+1}} = \sqrt{2}.\]
Therefore, the maximum minimum Euclidean distance of the column vectors in our code matrix is $\sqrt{2}$. Note that this Euclidean distance is not equivalent to the superimposed vectors minimum Euclidean distance.
\vspace{-0.2cm}
\section{Comparisons with other Algebraic LDS designs}
\label{simulation}
\begin{figure}[h]
\vspace{-.4cm}
	\centering
	\includegraphics[width=0.53\textwidth]{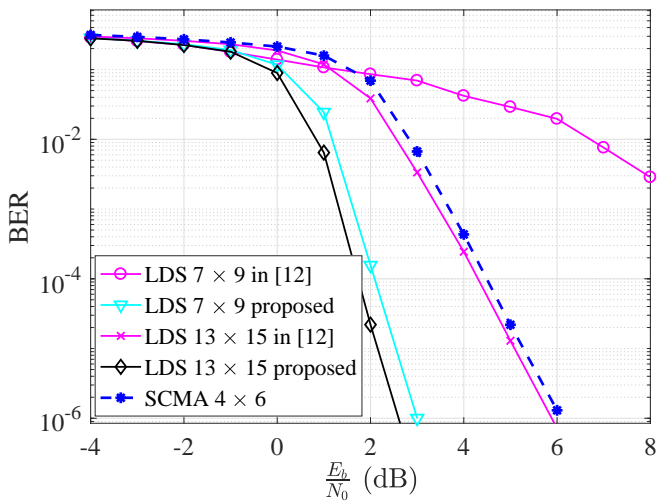}
	\caption{BER Performance of LDS and SCMA with Turbo Coding in Frequency-nonselective Rayleigh Fading.}\label{CodedRayleigh}
	\vspace{-.2cm}
\end{figure}

In this section, we assess the performance of our proposed LDS code sets against the LDS sets in \cite{Mheich2019}, for quadrature amplitude modulation (QAM) signaling. We observed that when the non-zero positions of the LDS code sets generated according to \cite{Mheich2019} are different than those of the proposed LDS code sets, the bit error rate (BER) performance of the former deteriorates substantially. Therefore, while generating the LDS codes of sizes $7 \times 9$ and $13 \times 15$ based on \cite{Mheich2019}, the non-zero entries are kept in the same positions as those in the proposed LDS sets; however, the values of the non-zero entries are obtained by following the exact procedure outlined in \cite{Mheich2019}.
\begin{figure}[h]
\vspace{-.4cm}
	\centering
	\includegraphics[width=0.53\textwidth]{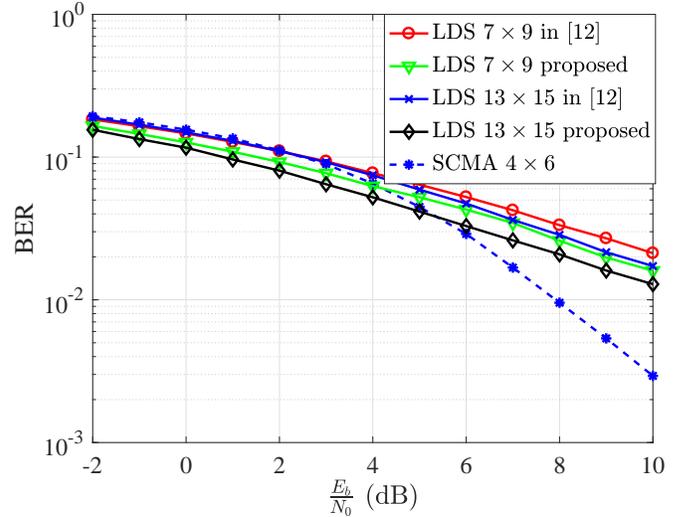}
	\caption{Uncoded BER performance of LDS and SCMA in frequency-nonselective Rayleigh fading.}\label{uncoded}
	\vspace{-.2cm}
\end{figure}
It should also be noted that, the proposed LDS code sets are not necessary uniquely decodable (UD) under QAM signaling; as such, the BER performance will have an error floor under the maximum-likelihood (ML) detection over the additive white Gaussian noise (AWGN) channel.
\begin{figure}[h]
\vspace{-.2cm}
	\centering
	\includegraphics[width=0.53\textwidth]{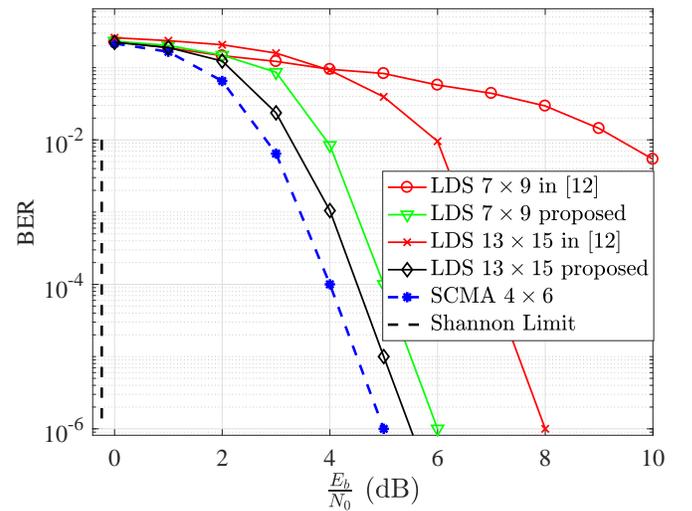}
	\caption{BER Performance of LDS and SCMA with Turbo Coding in AWGN.}\label{CodedAWGN}
	\vspace{-.1cm}
\end{figure}
Due to this reason we observed that the proposed LDS codes perform much better over the Rayleigh fading channels as opposed to AWGN channel as shown in Figs. \ref{uncoded} - \ref{CodedRayleigh}. In addition, the BER hinges not only on the minimum distance criterion, but also on the average Gaussian separability margin \cite{MichelHanzo2021}.
 Simulations were performed over the AWGN and Rayleigh fading channels with the fading rate of the symbol duration. We can apply transmitter precoding scheme for frequency selective channels \cite{MichelHanzo2021}. In our simulations, we utilized LDS spreading codes of sizes $7 \times 9$ and $13 \times 15$, SCMA \cite{Altera5G} of size $4 \times 6$ and the corresponding information rates in case of 4QAM are $\eta_{LDS}= c_r \cdot b_s\cdot 9/7=0.86 $, and $\eta_{LDS}=c_r \cdot b_s\cdot 15/13=0.77$ bits/s/Hz, where $b_s=2$ bits/symbol and $c_r=1/3$ code rate, respectively.
 Therefore, the corresponding unrestricted Shannon limits are calculated by using the upper bound $\log_2{(1+\gamma \beta)}$ as $E_b/N_o = (2^{\eta_{LDS}}-1)/\eta_{LDS}$, $E_b/No = 0.947 \:\:(-0.24\: \text{dB})$ and  $E_b/No = 0.916 \:\:(-0.38 \:\text{dB})$ for $\eta_{LDS}=0.86$ and $\eta_{LDS}=0.77$ where $\gamma$ denotes signal-to-noise ratios (SNR) and $\beta =K/L$ denotes the overload factor, respectively.
Fig. \ref{uncoded} shows that when no error control coding is used, SCMA performs better than LDS using our proposed LDS codes at high $E_b/N_0$ in Rayleigh fading. However, we also note that the BER performance of our proposed LDS scheme performs better at low $E_b/N_0$. When Turbo coding with the rate of $1/3$, generator polynomials of $1+x+x^2$, $1+x^2+x^3$, a feedback connection polynomial of $1+x+x^2$ and interleaver is used; however, our proposed LDS technique outperforms SCMA in Rayleigh fading as demonstrated in Fig. \ref{CodedRayleigh}.  This is because the energy per code bit to single sided noise spectral density ratio at the input to the decoder is low. At a BER of $10^{-3}$, LDS using our proposed spreading codes provides approximately a $3\:\text{dB}$ improvement over SCMA in Rayleigh fading. In all of our simulations, we used message passage algorithm (MPA) detector for SCMA and probabilistic data association (PDA) \cite{Pattipati2004} multiuser detector for all LDS codes. The reason we used PDA detector as its performance is similar to other best low-complexity detectors. In contrast to the PDA, the MPA does not need to perform any matrix inversion, but its complexity increases exponential by both with the size of the symbol alphabet $M$ and number of non-zero positions of the spreading waveform $d_f$. Fortunately matrix inversion required by PDA can be carried out quite efficiently with the aid of the Sherman–Morrison–Woodbury formula at an overall complexity order of $\mathcal{O}(K^3)$. 
\vspace{-.3cm}
\section{Conclusion}
\label{conclusion}
In this paper, we conceived an improved low-density spreading (LDS) sequence design based on an algebraic scheme. We developed a novel LDS construction based on projective geometry. In terms of its bit error rate (BER) performance, our proposed improved LDS code set outperforms the existing LDS designs over the frequency-nonselective Rayleigh fading and additive white Gaussian noise (AWGN) channels. We demonstrated that achieving the best BER depends on the minimum distance. Our future research will consider developing a theory for the overloaded cases that has greater number of users than the proposed construction. 

\bibliographystyle{IEEEtran}
\bibliography{IEEEabrv,michelbib_file}

\begin{thebibliography}{10}
\providecommand{\url}[1]{#1}
\csname url@samestyle\endcsname
\providecommand{\newblock}{\relax}
\providecommand{\bibinfo}[2]{#2}
\providecommand{\BIBentrySTDinterwordspacing}{\spaceskip=0pt\relax}
\providecommand{\BIBentryALTinterwordstretchfactor}{4}
\providecommand{\BIBentryALTinterwordspacing}{\spaceskip=\fontdimen2\font plus
\BIBentryALTinterwordstretchfactor\fontdimen3\font minus
  \fontdimen4\font\relax}
\providecommand{\BIBforeignlanguage}[2]{{%
\expandafter\ifx\csname l@#1\endcsname\relax
\typeout{** WARNING: IEEEtran.bst: No hyphenation pattern has been}%
\typeout{** loaded for the language `#1'. Using the pattern for}%
\typeout{** the default language instead.}%
\else
\language=\csname l@#1\endcsname
\fi
#2}}
\providecommand{\BIBdecl}{\relax}
\BIBdecl

\bibitem{Dai2018}
L.~{Dai}, B.~{Wang}, Z.~{Ding}, Z.~{Wang}, S.~Chen, and L.~{Hanzo}, ``A survey
  of non-orthogonal multiple access for 5{G},'' \emph{IEEE Commun. Surveys \&
  Tuts.}, vol.~20, no.~3, pp. 2294--2323, Thirdquarter 2018.

\bibitem{Hoshyar2008}
R.~{Hoshyar}, F.~P. {Wathan}, and R.~{Tafazolli}, ``Novel low-density signature
  for synchronous {CDMA} systems over {AWGN} channel,'' \emph{IEEE Trans.
  Signal Process.}, vol.~56, no.~4, pp. 1616--1626, Apr. 2008.

\bibitem{Razavi2012}
R.~{Razavi}, M.~{AL-Imari}, M.~A. {Imran}, R.~{Hoshyar}, and D.~{Chen}, ``On
  receiver design for uplink low density signature {OFDM (LDS-OFDM)},''
  \emph{IEEE Trans. Commun.}, vol.~60, no.~11, pp. 3499--3508, Nov. 2012.

\bibitem{Nikopour2013}
H.~{Nikopour} and H.~{Baligh}, ``{Sparse code multiple access},'' in
  \emph{Proc. IEEE Pers., Indoor, Mobile Radio Conf. (PIMRC)}, London, U.K.,
  Sep. 2013, pp. 332--336.

\bibitem{Hoshyar2006}
R.~{Hoshyar}, F.~P. {Wathan}, and R.~{Tafazolli}, ``Novel low-density signature
  structure for synchronous {DS-CDMA} systems,'' in \emph{Proc. IEEE Global
  Telecommun. Conf. (GLOBECOM)}, San Francisco, U.S.A., Nov. 2006, pp. 1--5.

\bibitem{Song2017}
G.~{Song}, X.~{Wang}, and J.~{Cheng}, ``Signature design of sparsely spread
  code division multiple access based on superposed constellation distance
  analysis,'' \emph{IEEE Access}, vol.~5, pp. 23\,809--23\,821, Oct. 2017.

\bibitem{Jiang2019}
K.~{Lu} and C.~{Jiang}, ``Optimized low density superposition modulation for
  {5G} mobile multimedia wireless networks,'' \emph{IEEE Access}, vol.~7, pp.
  174\,227--174\,235, Dec. 2019.

\bibitem{JVan2009}
J.~{van de Beek} and B.~M. {Popovic}, ``Multiple access with low-density
  signatures,'' in \emph{Proc. IEEE Global Telecommun. Conf. (GLOBECOM)},
  Honolulu, U.S.A., Nov. 2009, pp. 1--6.

\bibitem{Claude1995}
C.~{D'Amours} and A.~{Yongacoglu}, ``Hybrid {DS/FH-CDMA} system employing
  {MT-FSK} modulation for mobile radio,'' in \emph{Proc. IEEE Pers., Indoor,
  Mobile Radio Conf. (PIMRC)}, vol.~1, Toronto, Canada, Sep. 1995, pp.
  164--168.

\bibitem{Ivanov2013}
F.~I. {Ivanov} and V.~V. {Zyablov}, ``Low-density parity-check codes based on
  {S}teiner systems and permutation matrices,'' \emph{Probl. Inf. Transm.},
  vol.~49, p. 333–347, Oct. 2013.

\bibitem{Atkin2018}
Y.~{Wu}, E.~{Attang}, and G.~E. {Atkin}, ``A novel {NOMA} design based on
  steiner system,'' in \emph{Proc. IEEE Int. Conf. Electro/Inf. Technol.
  (EIT)}, Rochester, U.S.A., May 2018, pp. 0846--0850.

\bibitem{Mheich2019}
Z.~{Liu}, P.~{Xiao}, and Z.~{Mheich}, ``Power-imbalanced low-density signatures
  ({LDS}) from {Eisenstein} numbers,'' in \emph{Proc. IEEE VTS Asia Pacific
  Wireless Commun. Symp. (APWCS)}, Singapore, Aug. 2019, pp. 1--5.

\bibitem{Xudong2021}
X.~Li, Z.~Gao, Y.~Gui, Z.~Liu, P.~Xiao, and L.~Yu, ``Design of power-imbalanced
  {SCMA} codebook,'' \emph{IEEE Trans. on Vehic. Tech.}, pp. 1--5, Dec. 2021.

\bibitem{Singer1938}
J.~{Singer}, ``{A theorem of finite projective geometry and some applications
  to number theory},'' \emph{Transactions of the American Mathematical
  Society}, vol.~43, no.~3, pp. 377--385, May 1938.

\bibitem{Arasu1995}
K.~T. {Arasu}, J.~F. {Dillon}, D.~{Jungnickel}, and A.~{Pott}, ``The solution
  of the {Waterloo} problem,'' \emph{Journal of Combinatorial Theory Ser. A},
  vol.~71, no.~2, pp. 316--331, Aug. 1995.

\bibitem{MichelHanzo2021}
M.~{Kulhandjian}, H.~{Kulhandjian}, C.~{D'Amours}, and L.~{Hanzo},
  ``Low-density spreading code design based on {G}aussian separability,''
  \emph{IEEE Access}, vol.~9, pp. 33\,963--33\,993, Mar. 2021.

\bibitem{Altera5G}
``1st 5g algorithm innovation competition-envl.o-scma,''
  \url{http://www.innovateasia.com/5g/images/pdf/1st\%205G\%20Algorithm\%20Innovation\%20Competition-ENV1.0\%20-\%20SCMA.pdf},
  Altera Innovate Asia website.

\bibitem{Pattipati2004}
D.~{Pham}, K.~R. {Pattipati}, P.~K. {Willett}, and J.~{Luo}, ``{A generalized
  probabilistic data association detector for multiple antenna systems},''
  \emph{IEEE Commun. Lett.}, vol.~8, no.~4, pp. 205--207, Apr. 2004.

\end{thebibliography}
\end{document}